\documentclass[]{spie}  

\usepackage{amsmath,amsfonts,amssymb}
\usepackage{graphicx}
\usepackage[colorlinks=true, allcolors=blue]{hyperref}
\usepackage{booktabs}
\usepackage{siunitx}
\usepackage{microtype}

\DeclareSIUnit\arcsecond{as}
\DeclareSIUnit\microarcsecond{\micro\arcsecond}

\newcommand{\uas}{\si{\microarcsecond}}

\title{ADORA: a differentiable optical modeling and astrometric retrieval framework for SHERA}

\author[a]{Dylan McKeithen}
\author[a]{Stuart Shaklan}
\author[a]{Gautam Vasisht}
\author[a]{Joe Green}
\author[a]{Rhonda Morgan}
\author[b]{Louis Desdoigts}
\author[c]{Peter Tuthill}
\affil[a]{Jet Propulsion Laboratory, California Institute of Technology, Pasadena, California USA}
\affil[b]{Leiden University, Leiden, Netherlands}
\affil[c]{University of Sydney, Sydney, Australia}

\authorinfo{Further author information: (Send correspondence to Dylan McKeithen)\\Dylan McKeithen: E-mail: dylan.m.mckeithen@jpl.nasa.gov, Telephone: 1 818 393 3993}

\begin{document} 
\maketitle

\begin{abstract}
Searching for Habitable Exoplanets with Relative Astrometry (SHERA) is a proposed Small Explorer mission concept designed to measure the separation of nearby binary stars at microarcsecond-class precision. Recovering this signal requires separating astrophysical motion from coupled changes in pointing, plate scale, wavefront error, spectral response, and detector calibration.
We present the Astrometric Differentiable Optics and Retrieval Algorithm (ADORA), an image-domain framework that combines a three-plane differentiable physical-optics model with a layered astrometric inference algorithm. The forward model includes a diffractive pupil, mirror-specific wavefront error and beamwalk, polychromatic source and throughput models, and configurable detector effects.
Per-frame registration states are treated locally and eliminated through Schur reduction before the slower astrometric and instrument state is updated in a prior-whitened Fisher eigenbasis. 
Five-minute matched-model simulations show no detected separation bias at the current Monte Carlo depth and approximately $11~\uas$ realization-to-realization scatter. 
A SHERA target sweep reveals a more-than-fivefold variation in astrometric information between Alpha Centauri and 61 Cygni, motivating future target-dependent accumulation and update cadence. 
High-order-wavefront knowledge error can drive the retrieval toward a strongly biased astrometric solution while leaving the local posterior sigma nearly unchanged, demonstrating that statistical curvature alone does not capture unmodeled bias. 
Pixel-position errors across the tested range remain near the matched-model recovery scale indicating robustness to certain detector calibration errors.
ADORA provides a flexible framework for studying astrometric extraction, calibration-bias diagnosis, and future SHERA requirements.

\end{abstract}


\section{INTRODUCTION}
\label{sec:intro}

Detecting the astrometric motion induced by terrestrial planets around nearby Sun-like stars requires sensitivity to angular perturbations at the microarcsecond scale. Narrow-angle relative astrometry offers a particularly favorable geometry when both components of a nearby binary are bright: the companion provides an in-field reference, the small field suppresses differential optical distortion, and the measurement avoids the photon-noise penalty associated with faint background reference stars \cite{Shao1992,Lane2004}. Repeated measurements of the projected separation between the binary components can then reveal a periodic planet-induced perturbation superposed on the much larger binary orbit. The full scientific case for applying this method to nearby Sun-like binaries is developed in a companion SHERA mission study \cite{Christiansen2026}; the present paper focuses on the instrument-facing problem of recovering the relative-astrometric measurement from a high-cadence image sequence.

Searching for Habitable Exoplanets with Relative Astrometry (SHERA) is a proposed Small Explorer mission concept built around a compact, \SI{22}{\centi\meter} F/35 Ritchey–Chrétien telescope with a reflective diffractive pupil etched onto the primary mirror. The mission would repeatedly observe a set of bright nearby binary systems over a multi-year baseline, using the structured PSF produced by the diffractive-pupil as a self-calibrating metrology signal \cite{Christiansen2026}. High target flux supports rapid short-exposure imaging, while the narrow field reduces differential distortion and permits a comparatively simple optical architecture. The diffractive pupil spreads the stellar light into information-rich features that retain astrometric sensitivity while also encoding changes in the optical state. The central challenge is to extract the astrometric signal from the short-exposure image sequence while separating true relative motion from changes in the instrument.

At the required precision, changes in the instrument state including plate-scale drift, mirror-specific wavefront error (WFE), pointing-dependent beamwalk, effective-wavelength variation, detector geometry and response, or exposure-integrated jitter and smear, can mask the true astrometric motion induced by a potential exoplanet. The diffractive pupil provides sensitivity to these instrumental degrees of freedom, but exploiting that information requires a physical image model that can be evaluated and differentiated jointly with respect to the science and calibration parameters.

We therefore present the Astrometric Differentiable Optics and Retrieval Algorithm (ADORA), a forward-modeling and observation-level inference framework for SHERA. ADORA is built on the \texttt{dLux}/JAX differentiable-optics stack \cite{Desdoigts2023,Desdoigts2024} and explicitly models propagation from the primary mirror to the secondary mirror and focal plane, including the diffractive pupil, mirror-specific low- and high-order wavefront error, beamwalk, target spectra, throughput, and configurable detector effects. The associated retrieval separates rapidly varying  ``fast'' image-registration parameters from the ``slow'' astrometric and instrument state. Registration variables are solved locally in short sub-blocks, their contribution is eliminated through a Schur complement, and the remaining slow-state information is combined within each update window. The proposed correction is then regularized in a prior-whitened Fisher information eigenbasis before the next window is processed. The final output is an observation-level separation estimate and a set of calibration and fit-quality diagnostics for downstream multi-epoch orbit analysis.

This paper makes three principal contributions. First, it develops a three-plane differentiable physical-optics model tailored to the SHERA measurement concept. Second, it formulates a layered image-domain retrieval that avoids one monolithic optimization over every image, registration variable, and calibration parameter. Third, it introduces a controlled truth--inference validation framework for separating stochastic retrieval behavior from systematic calibration bias. Initial campaigns establish matched-model recovery and show that a static high-order wavefront calibration error can drive the estimator toward a repeatable biased solution. These results are an initial validation and sensitivity study rather than a complete SHERA error budget or final mission-performance prediction.

Section~\ref{sec:overview} defines the SHERA observable and the ADORA parameter hierarchy. Section~\ref{sec:forward-model} describes the differentiable three-plane image-formation model, and Sec.~\ref{sec:retrieval-algorithm} presents the layered retrieval algorithm. Section~\ref{sec:validation-design} defines the validation campaigns and reporting conventions. Results are presented in Sec.~\ref{sec:results}, with interpretation, limitations, and future priorities discussed in Sec.~\ref{sec:discussion}.

\section{SHERA Measurement Concept and ADORA Overview}
\label{sec:overview}

SHERA plans to observe each science target for approximately 30 minutes at a time before re-targeting, and will acquire frames at \SI{20}{\hertz} producing a long sequence of short-exposure focal-plane images.  
The goal of the ADORA analysis chain is to reduce this sequence of images into a calibrated astrometric measurement for each science-observation epoch. 

ADORA organizes the stream of data into three nested levels. Individual \SI{50}{\milli\second} frames retain the instantaneous image and registration information. Groups of nominally 20 frames form \SI{1}{\second} sub-blocks, within which the local registration state is solved. Multiple sub-blocks are combined into block updates, nominally of order \SI{30}{\second}, over which the observation-level state is assumed to remain fixed. The precise partition and update cadence is configurable and can be adapted to target brightness, pointing behavior, and computational constraints.

\subsection{Relative astrometric observable}
\label{sec:relative-observable}

The fundamental science observable considered here is the sky-projected separation magnitude between the two components of a target binary. If $\Delta\boldsymbol{r}=(\Delta x,\Delta y)$ is the projected relative-position vector, the scalar separation is

\begin{equation}
    \rho = \left\lVert \Delta\boldsymbol{r} \right\rVert
         = \sqrt{\Delta x^2+\Delta y^2}.
    \label{eq:binary-separation}
\end{equation}

SHERA repeatedly measures this quantity over the mission, with planet-induced motion appearing as a small perturbation to the otherwise dominant binary orbit. The current ADORA validation framework estimates $\rho$ independently for each nominally \SI{30}{\minute} science observation. The resulting observation-epoch estimate is accompanied by a covariance or posterior uncertainty, fit-quality diagnostics, and the calibration and processing context needed to interpret the measurement downstream.

The per-observation separation estimates are passed to MARA, the Microarcsecond Astrometry Retrieval Algorithm \cite{Roberson2026}, which performs the multi-epoch orbit-fitting and planet-detection analysis to detect, constrain or otherwise rule out candidate exoplanets. ADORA therefore occupies the instrument-facing portion of the mission analysis chain and produces a time series of astrometric measurements used for planet searches.

Recovering the separation requires jointly accounting for several quantities that influence the focal-plane image. In the current baseline retrieval algorithm, ADORA fits the total source flux and binary contrast together with the instrument plate scale and low-order Zernike wavefront-error coefficients (Noll indices 4-11) on both the primary (M1) and secondary (M2) mirrors. Short-timescale image translation and roll are solved locally from individual frames. Other quantities---including the component spectral energy distributions, high-order wavefront maps, filter throughput, detector quantum efficiency, pixel positions, and pixel response---enter as calibration inputs to the forward model. These calibration quantities are held fixed in the baseline retrieval but may differ between the truth and inference models to enable controlled knowledge-error studies.

\subsection{Parameter hierarchy: science, nuisance, and calibration states}
\label{sec:parameter-hierarchy}

ADORA organizes the inference according to the characteristic timescales and roles of the model parameters. We denote the observation-level fitted state by $\boldsymbol{\theta}$. For the baseline binary retrieval, this state can be written schematically as

\begin{equation}
    \boldsymbol{\theta}
    =
    \left[
        \rho,\,
        \log F_{\mathrm{tot}},\,
        C,\,
        s,\,
        \boldsymbol{a}^{\mathrm{M1}}_{\mathrm{LO}},\,
        \boldsymbol{a}^{\mathrm{M2}}_{\mathrm{LO}}
    \right]^{\mathsf{T}},
    \label{eq:slow-state}
\end{equation}

where $F_{\mathrm{tot}}$ is the total detected flux, $C$ is the component contrast or flux ratio, $s$ is the plate scale, and $\boldsymbol{a}^{\mathrm{M1}}_{\mathrm{LO}}$ and $\boldsymbol{a}^{\mathrm{M2}}_{\mathrm{LO}}$ are the active low-order wavefront coefficients. The scalar separation, $\rho$, is the target science product, but the remaining terms must be estimated jointly because errors in these quantities can produce apparent astrometric motion.

The local nuisance state for sub-block $b$ is denoted $\boldsymbol{\phi}_b$ and contains the frame-dependent registration parameters, nominally the horizontal and vertical translation and roll of each frame. These variables are required to explain the image sequence, but they are not retained as observation-level science products. ADORA instead solves them within each short sub-block and eliminates their local contribution before information is accumulated into the slower state.

A third group consists of calibration quantities, denoted collectively by $\boldsymbol{\kappa}$. These describe the instrument and source information supplied to the image model, including component spectra, high-order wavefront maps, bandpass and detector quantum efficiency, and detector pixel-position and pixel-response calibrations. They are generally fixed or tightly constrained during routine astrometric extraction. Treating them as a distinct calibration state makes the truth--inference relationship explicit and provides a systematic way to inject controlled model mismatch. Parameters in $\boldsymbol{\kappa}$ may also be promoted into the fitted state in future calibration modes when the available data provide sufficient information.

\begin{table}[htbp]
\caption{Parameter hierarchy used by the current ADORA retrieval. The fitted slow state contains the separation together with the coupled source and instrument terms.}
\label{tab:parameter-groups}
\centering
\small
\begin{tabular}{@{}p{0.17\linewidth}p{0.30\linewidth}
                p{0.17\linewidth}p{0.28\linewidth}@{}}
\toprule
Parameter group & Representative quantities & Characteristic timescale & Role in the current retrieval \\
\midrule

Local nuisance state, $\boldsymbol{\phi}_b$
& Horizontal and vertical registration and image roll
& Per-frame
& Solved locally in each sub-block and marginalized before slow-state
information is accumulated. \\

Observation-level state, $\boldsymbol{\theta}$
& Binary separation, total flux, contrast, plate scale, and low-order wavefront error on M1 + M2
& Constant within an update block; updated across the science observation
& Jointly estimated slow state. Separation is the science product; the remaining terms control coupled source and instrument effects. \\

Calibration state, $\boldsymbol{\kappa}$
& Source spectra, high-order M1/M2 wavefront maps, filter response, detector quantum efficiency, pixel positions, and pixel response
& Established through dedicated calibration; treated as static
& Perturbed between truth and inference models in knowledge-error studies. \\
\bottomrule
\end{tabular}
\end{table}

This hierarchy avoids two undesirable extremes: fitting every uncertain instrument quantity independently in every frame, and assuming that the instrument is perfectly known and static. Instead, fast registration states are handled locally, the slow fitted state is updated using information accumulated over the observation, and externally calibrated quantities are included with an explicitly defined level of knowledge.

\subsection{Observation-level information flow}
\label{sec:observation-flow}

For each sub-block, the forward model is fitted directly to the focal-plane images while the current slow reference state is held fixed. The frame-dependent registration variables are eliminated through a Schur complement, leaving a reduced score and information matrix that summarize what the sub-block constrains about $\boldsymbol{\theta}$ after the nuisance state has been adjusted. Summaries sharing a common reference are accumulated within an update window, and the resulting slow-state correction is regularized in a Fisher information eigenbasis before becoming the reference for the next window. Conditional on the shared state, the sub-block solves are independent and can be executed in parallel; Sec.~\ref{sec:retrieval-algorithm} gives the formal construction.


This architecture is related to local-variable elimination in bundle adjustment and factor-graph state estimation \cite{Triggs2000,DellaertKaess2017}, as well as structured global astrometric solutions such as the Gaia astrometric core solution \cite{Lindegren2012}. ADORA adapts these ideas to a differentiable, image-domain physical-optics likelihood, preserving focal-plane information without an observation-wide fit over every frame and nuisance parameter.


\section{Differentiable Forward Model}
\label{sec:forward-model}

ADORA uses a differentiable image-formation model to connect the astrophysical scene and instrument state to the expected focal-plane data. The implementation is built on the \texttt{dLux} physical-optics framework, which uses JAX automatic differentiation to propagate derivatives through optical and detector models \cite{Desdoigts2023}. Related applications of \texttt{dLux} have demonstrated its use for Fisher-information-based optical design and calibration of interferometric data \cite{Desdoigts2026}. For SHERA, this framework is extended to include target-dependent spectral weighting, explicit Fresnel propagation between the primary and secondary mirrors allowing for mirror-specific wavefront error, and a configurable detector model.

For a frame acquired at time $t_f$, the model predicts a detector expectation $\boldsymbol{\mu}_f$ from a source state, pointing state, optical state, and calibration state. Schematically,

\begin{equation}
    \boldsymbol{\mu}_f
    =
    \mathcal{D}_{\boldsymbol{\kappa}_{\mathrm{det}}}
    \left[
        \sum_{c=1}^{N_{\mathrm{src}}}
        \sum_{k=1}^{N_{\lambda}}
        w_{c,k}\,
        \left|
            \mathcal{H}_{\lambda_k}
            \left(
                \boldsymbol{\alpha}_{c,f},
                \boldsymbol{\kappa}_{\mathrm{opt}}
            \right)
        \right|^2
    \right],
    \label{eq:forward-model}
\end{equation}

where $c$ indexes source components, $\lambda_k$ indexes the wavelength grid, $w_{c,k}$ is the detected spectral weight of source $c$, and $\mathcal{H}_{\lambda_k}$ denotes coherent propagation through the telescope. The detector operator $\mathcal{D}$ applies the configured sampling, intra-pixel, diffusion, pointing-blur, and calibration effects. Image noise is applied after evaluation of this deterministic expectation when synthetic observations are generated. Because the operators in Eq.~(\ref{eq:forward-model}) are differentiable, ADORA can evaluate image sensitivities with respect to source, registration, and instrument parameters without constructing a separate empirical inverse model.

\subsection{Target and scene model}
\label{sec:target-scene-model}

The astrophysical scene model supplies the source-related inputs to the instrument simulation. For a binary target, these inputs include the nominal component separation and position angle, the field registration of the system, total detected flux, component contrast, and a spectral energy distribution for each star. The component locations can be written as angular offsets relative to a common field reference. At frame $f$, the apparent angle of source component $c$ is represented schematically by

\begin{equation}
    \boldsymbol{\alpha}_{c,f}
    =
    \boldsymbol{\alpha}_{c,0}
    \left(
        \rho,\psi,\boldsymbol{x}_0
    \right)
    +
    \boldsymbol{p}(t_f),
    \label{eq:source-pointing}
\end{equation}

where $\rho$ and $\psi$ describe the binary geometry, $\boldsymbol{x}_0$ is the nominal field registration, and $\boldsymbol{p}(t_f)$ is the time-dependent spacecraft pointing state. This separation between source geometry and common pointing motion allows the same binary scene to be rendered along an arbitrary trajectory while preserving the relative astrometric observable.

The current baseline uses Alpha Centauri A/B, but targets are selectable through a registry in the model. Each entry supplies the nominal astrometry, photometry, component spectra, and observing geometry needed to instantiate the scene. The same instrument forward model can then be applied for the broader list of SHERA targets, including systems with lower flux, different contrast, and different angular separation. The scene abstraction can also accommodate additional sources or background structure, although the current validation campaigns consider only the two-component binary scene.

Observation timing and trajectory modeling is part of the scene definition. Each frame is assigned an exposure interval and a time-dependent pointing state. For trajectory-driven simulations, ADORA reads a continuous spacecraft trajectory, applies any configured preprocessing like high-pass filtering, interpolates it to the exposure cadence, and derives the registration and smear quantities used by the renderer. The grouping of frames into sub-blocks and update blocks is an inference choice described in Sec.~\ref{sec:retrieval-algorithm}; the forward model itself operates at the individual-frame level.

\subsection{Three-plane optical propagation}
\label{sec:three-plane-propagation}

The SHERA optical model explicitly represents the primary-mirror plane, the secondary-mirror plane, and the focal plane. At the primary, the complex field for source component $c$ and wavelength $\lambda$ is formed from the entrance aperture, the diffractive-pupil optical-path-difference map, the primary-mirror wavefront error, and the phase tilt associated with the apparent source angle:

\begin{equation}
    E_{\mathrm{M1}}(\boldsymbol{r};\lambda)
    =
    A_{\mathrm{M1}}(\boldsymbol{r})
    \exp\left[
        \frac{2\pi i}{\lambda}
        \left(
            W_{\mathrm{DP}}(\boldsymbol{r})
            +
            W_{\mathrm{M1}}(\boldsymbol{r})
            +
            \boldsymbol{\alpha}_{c,f}\cdot\boldsymbol{r}
        \right)
    \right].
    \label{eq:m1-field}
\end{equation}

The field is propagated over the M1--M2 separation using a Fresnel angular spectrum operator. The secondary aperture and secondary-mirror wavefront error are then applied in the propagated plane. The resulting field is back-propagated to an exit-pupil sampling conjugate to the primary plane and transformed to the requested focal-plane sampling with a matrix Fourier transform. The monochromatic calculation is evaluated over the wavelength grid and incoherently summed according to the component-specific spectral weights.

Explicit propagation to M2 is required because primary- and secondary-mirror aberrations are not generally interchangeable once off-axis propagation and beamwalk are included. The two components of the binary arrive from different field angles and therefore do not sample the secondary identically. Changes in spacecraft pointing additionally translate the illumination pattern across the secondary mirror. Spatial structure in the M2 wavefront can consequently produce differential changes in the two point-spread functions, including apparent image displacement that can be confused with a change in binary separation. A single-pupil Fraunhofer model cannot represent this pointing-dependent sampling of mirror-specific error. The three-plane model is therefore central to evaluating how pointing and wavefront knowledge propagate into the astrometric measurement.

\subsection{Diffractive pupil and mirror-specific wavefront error}
\label{sec:diffractive-pupil-wfe}

The primary pupil contains a phase-only diffractive pattern that produces the structured focal-plane response used for SHERA astrometry. The current implementation uses a scaled diffractive-pupil design derived from the \texttt{dLux} astrometric modeling framework \cite{Desdoigts2024}. Its structured central point-spread function encodes information about source position and optical state that would not be retained by a centroid-only measurement. The pupil prescription also includes a special high-spatial-frequency spectral phase grating, which diffracts a portion of the light into spectral features in the outer field. These features provide an additional observable for constraining effective spectral shape and chromatic scale.

Wavefront error on each mirror is divided into low- and high-spatial-frequency components,

\begin{equation}
    W_j(\boldsymbol{r})
    =
    \sum_{n\in\mathcal{Z}} a_{j,n} Z_n(\boldsymbol{r})
    +
    W_{j,\mathrm{HO}}(\boldsymbol{r}),
    \qquad
    j\in\{\mathrm{M1},\mathrm{M2}\},
    \label{eq:wfe-decomposition}
\end{equation}

where $Z_n$ are Zernike modes and $W_{j,\mathrm{HO}}$ is a sampled high-order wavefront map. The current binary retrieval activates Noll indices 4--11 on each mirror as observation-level fitted parameters. High-order maps are supplied as fixed calibration inputs and may be generated synthetically or loaded from measured surface data. In the present full-fidelity campaign construction, the active low-order projections are removed from the high-order maps so that the two representations remain distinct.

ADORA maintains separate truth and inference descriptions of the optical state. High-order wavefront error may be present in both models, while a controlled residual

\begin{equation}
    \Delta W_{j,\mathrm{HO}}
    =
    W_{j,\mathrm{HO}}^{\mathrm{inf}}
    -
    W_{j,\mathrm{HO}}^{\mathrm{truth}}
    \label{eq:wfe-knowledge-error}
\end{equation}

defines the high-order wavefront knowledge error (KE). This construction separates the physical wavefront present in the simulated telescope from imperfect knowledge of that wavefront in the retrieval model. The low-order coefficients remain free to adjust during inference, allowing the validation campaigns to measure how unmodeled high-order structure is absorbed by the fitted low-order state or leaks into the recovered astrometry.

\subsection{Spectral throughput model}
\label{sec:spectral-throughput}

The monochromatic point-spread function changes with wavelength through both diffraction and the phase-to-wave conversion of each optical-path-difference map. Accurate polychromatic rendering therefore requires a detected spectrum for each binary component. ADORA constructs this spectrum from the component stellar spectral energy distribution, the M2 bandpass response, and the detector quantum efficiency. On a discrete wavelength grid, the component weights are proportional to

\begin{equation}
    w_{c,k}
    \propto
    S_c(\lambda_k)\,
    T_{\mathrm{M2}}(\lambda_k)\,
    Q_{\mathrm{det}}(\lambda_k)\,
    \Delta\lambda_k ,
    \label{eq:spectral-weight}
\end{equation}

with any additional configured throughput terms incorporated in the same product. The weights are normalized separately from the fitted total-flux and contrast parameters so that a change in spectral shape does not silently redefine the source photometry.

The spectral model is target aware: the two stars may use distinct component spectra while sharing the same optical and detector response curves. ADORA also supports separate truth and inference spectral states, including truncation, response-curve perturbation, and out-of-band handling. In the baseline retrieval, the spectral quantities are fixed calibration inputs rather than active observation-level parameters. Spectral mismatch has yet to be studied in depth, but remains an important area of focus.

\subsection{Detector, jitter, smear, and noise model}
\label{sec:detector-model}

After optical propagation and polychromatic summation, the focal-plane intensity is passed through a configurable detector-layer stack. The current model includes finite pixel modulation transfer, charge diffusion, pixel-position calibration, and pixel-response or flat-field calibration. Pixel modulation and charge diffusion are represented through convolutional response layers, while pixel-position and pixel-response maps encode geometric and radiometric departures from an ideal detector. As with the optical and spectral states, separate truth and inference detector maps can be used to inject controlled calibration knowledge error.

Pointing effects that are not represented solely by the instantaneous source registration can be included as image-domain blur. An optional jitter layer represents unresolved high-frequency motion through a compact convolution kernel. Exposure-time smear is derived from the change in the pointing trajectory during an integration and can be represented by a frame- or sub-block-specific kernel. In the current trajectory-driven campaigns, the smear applied to the inference model can be matched to the rendered trajectory or deliberately mismatched to test sensitivity to pointing knowledge.

Synthetic observations are generated by drawing from the deterministic expectation after all configured detector layers have been applied. The baseline noise model includes photon-counting noise and additive detector read noise; dark current and other terms can be enabled when required. The associated image variance is retained for use in the image likelihood and fit-quality diagnostics. Separating the expected image from its stochastic realization permits repeated noise draws from the same physical state and makes it possible to distinguish random measurement scatter from bias caused by truth--inference model mismatch.

Table~\ref{tab:forward-model-components} summarizes the principal forward-model components and their treatment in the current validation framework.

\begin{table}[htbp]
\caption{Principal components of the ADORA forward model. ``Fitted'' denotes quantities included in the current slow or local inference state; ``calibrated'' denotes quantities supplied to the model and optionally perturbed in knowledge-error studies.}
\label{tab:forward-model-components}
\centering
\footnotesize
\begin{tabular}{@{}p{0.18\linewidth}p{0.31\linewidth}
                p{0.20\linewidth}p{0.23\linewidth}@{}}
\toprule
Component & Physical role & Baseline treatment & Validation use \\
\midrule
Binary scene
& Separation, orientation, registration, photometry, and component spectra
& Separation and photometry fitted; spectra calibrated
& Target-registry, photometric, and spectral-mismatch studies \\

Pointing trajectory
& Frame registration, secondary-mirror beamwalk, jitter, and smear
& Registration fitted locally; trajectory supplies rendered motion
& Pointing capture, temporal-model, and smear-knowledge studies \\

Primary pupil
& Aperture, diffractive phase pattern, spectral grating, and M1 wavefront error
& Low-order WFE fitted; pupil and high-order WFE calibrated
& Diffractive-pupil design and M1 WFE knowledge-error studies \\

Secondary plane
& Fresnel-propagated illumination, M2 aperture, bandpass, and M2 wavefront error
& Low-order WFE fitted; high-order WFE and bandpass calibrated
& Beamwalk and M2 WFE knowledge-error studies \\

Spectral response
& Component SEDs, filter throughput, detector QE, and polychromatic weighting
& Fixed component-specific effective spectra
& Truth--inference effective-wavelength and response mismatch \\

Detector response
& Pixel MTF, charge diffusion, pixel positions, and pixel response
& Calibrated detector layers
& Pixel-position and flat-field knowledge-error studies \\

Pointing blur
& Unresolved jitter and finite-exposure trajectory smear
& Configured convolution or trajectory-derived kernels
& Jitter, smear, and pointing-knowledge sensitivity \\

Image noise
& Photon statistics and detector read noise
& Drawn for each synthetic observation
& Monte Carlo precision, likelihood, and residual validation \\
\bottomrule
\end{tabular}
\end{table}

\begin{figure}[htbp]
\centering
\includegraphics[width=\linewidth]{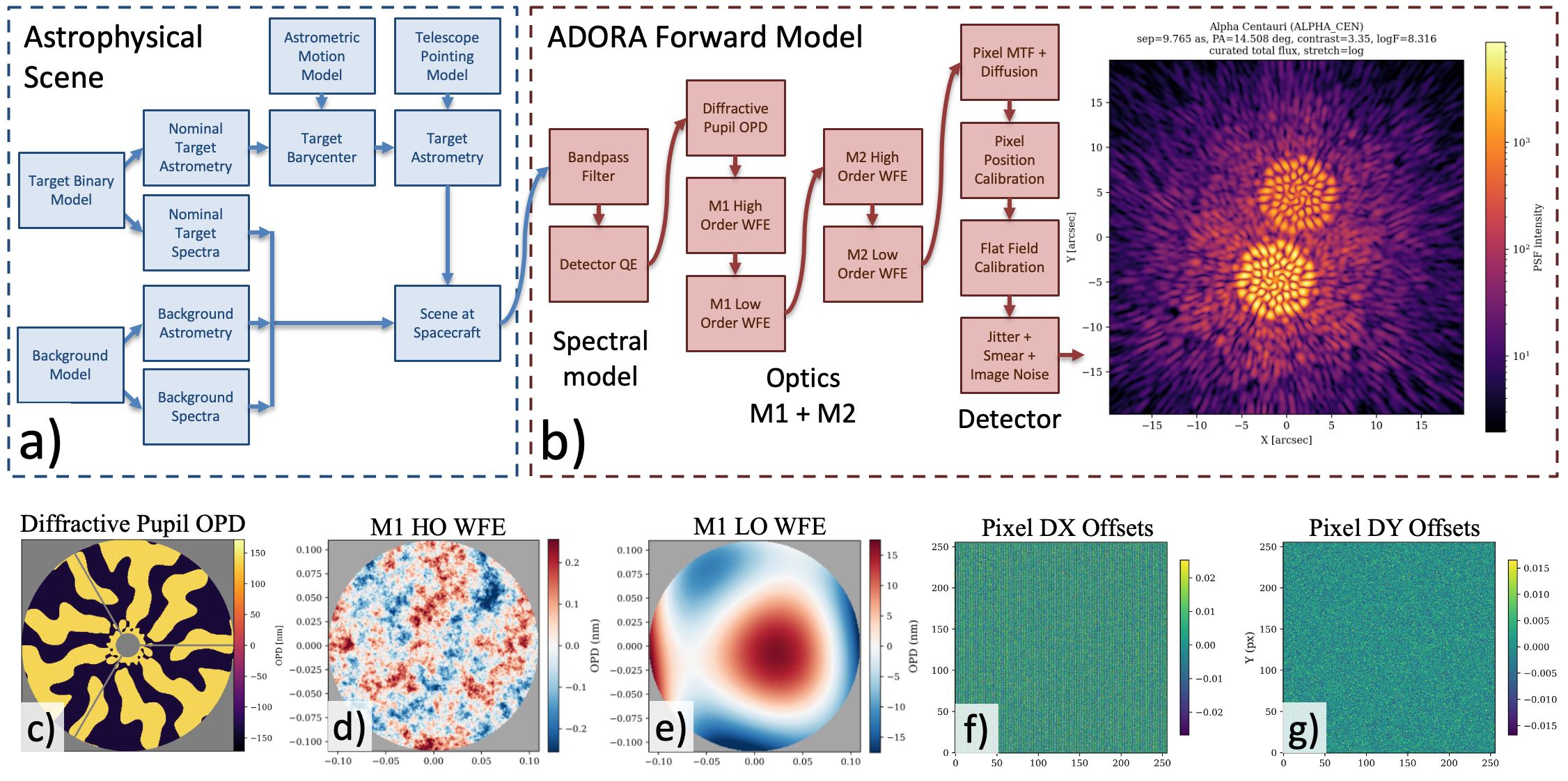}
\caption{Principal ingredients of the ADORA image-formation model. 
(a) The scene model combines target and background sources with astrometric motion and time-dependent X/Y/roll pointing, including exposure smear.
(b) Component spectra are weighted by the bandpass and detector QE, propagated through the diffractive M1 pupil and the M2 plane with mirror-specific wavefront error, and processed by the detector model to produce a focal-plane image.
(c)--(e) Example diffractive-pupil OPD, high-order WFE, and fitted low-order Zernike WFE on M1; analogous WFE terms are used on M2.
(f)--(g) Example detector-pixel X and Y position-offset maps. The detector chain also supports finite-pixel response, charge diffusion, flat-field calibration, jitter, smear, and image noise.
}
\label{fig:adora-forward-model}
\end{figure}

\section{Layered Astrometric Retrieval Algorithm}
\label{sec:retrieval-algorithm}

ADORA converts a high-cadence image sequence into one calibrated observation-epoch astrometric estimate without performing a single monolithic optimization over every image and every uncertain instrument parameter. A nominal \SI{30}{\minute} science observation is divided into sequential update blocks, each of which is further divided into short sub-blocks. The observation-level source and instrument state is assumed to be constant within an update block, while the image registration is allowed to vary on the frame timescale. This partition permits the local registration problems to be solved independently and in parallel, after which their information about the slower state is combined.

The nominal cadence used in the current validation campaigns is \SI{50}{\milli\second} per frame, 20 frames per approximately \SI{1}{\second} sub-block, and 30 sub-blocks per approximately \SI{30}{\second} update block. These sub-block and update-block durations can be adapted to the target flux, pointing behavior, instrument-stability timescale, and available computational resources.

\subsection{Image likelihood and parameter groups}
\label{sec:image-likelihood}

Let $\boldsymbol{d}_{b,f}$ denote the observed detector image in frame $f$ of sub-block $b$, and let $\boldsymbol{\mu}_{b,f}(\boldsymbol{\theta},\boldsymbol{\phi}_b)$ be the corresponding prediction from the differentiable forward model described in Sec.~\ref{sec:forward-model}. The observation-level state $\boldsymbol{\theta}$ contains the binary separation together with the photometric and instrumental quantities that must be estimated jointly. The local state $\boldsymbol{\phi}_b$ contains the frame-dependent registration parameters for sub-block $b$. For a diagonal image-variance model, the sub-block negative log likelihood can be written as

\begin{equation}
\begin{split}
    \mathcal{L}_b(\boldsymbol{\theta},\boldsymbol{\phi}_b)
    = \frac{1}{2}
    \sum_{f,p}
    \left[
        \frac{
            \left(
                d_{b,f,p}
                -
                \mu_{b,f,p}(\boldsymbol{\theta},\boldsymbol{\phi}_b)
            \right)^2
        }{
            v_{b,f,p}
        }
        +
        \ln v_{b,f,p}
    \right]
    +
    \mathcal{R}_{\theta}
    +
    \mathcal{R}_{\phi},
    \label{eq:subblock-likelihood}
\end{split}
\end{equation}

where $p$ indexes detector pixels, $v_{b,f,p}$ is the adopted variance, and $\mathcal{R}_{\theta}$ and $\mathcal{R}_{\phi}$ represent prior or temporal regularization terms when active. Constants independent of the fitted parameters have been omitted. The same formulation can accommodate a more general covariance model, although the present image-backed campaigns use a pixelwise variance representation.

The baseline observation-level state contains the parameters introduced in Eq.~(\ref{eq:slow-state}): binary separation, total flux, component contrast, plate scale, and low-order wavefront coefficients on M1 and M2. The nominal local state contains horizontal and vertical image registration and roll for each frame. Spectral and detector calibration quantities are fixed in the baseline solve, but the parameter partition is configurable; calibration quantities can be promoted into the fitted state when a dedicated observing mode provides enough information to constrain them.

\begin{table}[htbp]
\caption{Parameters used in the current baseline ADORA inference. Exact prior uncertainties and initialization amplitudes are specified with the validation configuration in Sec.~\ref{sec:validation-design}.}
\label{tab:inference-parameters}
\centering
\footnotesize
\begin{tabular}{@{}p{0.20\linewidth}p{0.15\linewidth}
                p{0.13\linewidth}p{0.17\linewidth}p{0.26\linewidth}@{}}
\toprule
Quantity & Symbol & Unit & Inference group & Current role \\
\midrule
Binary separation
& $\rho$
& arcsec
& Slow
& Target science quantity; updated between observation blocks. \\

Total detected flux
& $\log F_{\mathrm{tot}}$
& log photons
& Slow
& Jointly fitted source normalization. \\

Binary contrast
& $C$
& dimensionless
& Slow
& Jointly fitted component flux ratio. \\

Plate scale
& $s$
& arcsec pixel$^{-1}$
& Slow
& Jointly fitted instrumental scale. \\

M1 low-order WFE
& $\boldsymbol{a}^{\mathrm{M1}}_{\mathrm{LO}}$
& nm
& Slow
& Zernike coefficients, nominally Noll indices 4--11. \\

M2 low-order WFE
& $\boldsymbol{a}^{\mathrm{M2}}_{\mathrm{LO}}$
& nm
& Slow
& Zernike coefficients, nominally Noll indices 4--11. \\

Frame translation
& $x_{b,f},y_{b,f}$
& arcsec
& Local nuisance
& Solved within each sub-block and eliminated before accumulation. \\

Frame roll
& $\gamma_{b,f}$
& deg
& Local nuisance
& Activated where required by the binary registration model. \\
\bottomrule
\end{tabular}
\end{table}

\subsection{Sub-block registration solve}
\label{sec:subblock-registration}

For each sub-block, ADORA holds the slow state at a block-specific reference value $\boldsymbol{\theta}_{\mathrm{ref}}$ and solves

\begin{equation}
    \boldsymbol{\phi}_b^{\star}
    =
    \underset{\boldsymbol{\phi}_b}{\operatorname{arg\,min}}\;
    \mathcal{L}_b
    \left(
        \boldsymbol{\theta}_{\mathrm{ref}},
        \boldsymbol{\phi}_b
    \right).
    \label{eq:local-registration-solve}
\end{equation}

Automatic differentiation through the image model supplies the gradients required by the local optimizer. The current baseline uses only frame-local registration variables, while the implementation permits the addition of temporal terms such as drifting plate scale or WFE terms. The temporal model can therefore be matched to the available pointing estimate and to the timescale on which residual jitter is expected to vary.

The local solve absorbs the high-cadence pointing and registration motion before the slower source and instrument information is accumulated. It also produces diagnostics that are carried with the sub-block summary, including the fitted image residuals, reduced chi-square or related fit-quality statistics, optimizer convergence information, and registration histories. Sub-blocks or individual frames with poor fits can be retained with a warning, masked from the information summary, or rejected according to the configured policy. These checks are important because the curvature compression described below is meaningful only when the local solution and its neighborhood provide an adequate approximation to the image likelihood.

\subsection{Schur-reduced slow-state summaries}
\label{sec:schur-summaries}

Although the local optimization varies only $\boldsymbol{\phi}_b$, the fitted image sequence contains information about both the nuisance state and the slow state. ADORA evaluates the local gradient and curvature of the summed image likelihood around $(\boldsymbol{\theta}_{\mathrm{ref}},\boldsymbol{\phi}_b^{\star})$. For increments $\delta\boldsymbol{\theta}$ and $\delta\boldsymbol{\phi}_b$, the local quadratic approximation is

\begin{equation}
\begin{split}
    \mathcal{L}_b
    \simeq
    \mathcal{L}_{b,0}
    &+
    \begin{bmatrix}
        \boldsymbol{g}_{\theta,b} \\
        \boldsymbol{g}_{\phi,b}
    \end{bmatrix}^{\mathsf{T}}
    \begin{bmatrix}
        \delta\boldsymbol{\theta} \\
        \delta\boldsymbol{\phi}_b
    \end{bmatrix}
\\
    &+
    \frac{1}{2}
    \begin{bmatrix}
        \delta\boldsymbol{\theta} \\
        \delta\boldsymbol{\phi}_b
    \end{bmatrix}^{\mathsf{T}}
    \begin{bmatrix}
        \boldsymbol{H}_{\theta\theta,b}
        &
        \boldsymbol{H}_{\theta\phi,b}
        \\
        \boldsymbol{H}_{\phi\theta,b}
        &
        \boldsymbol{H}_{\phi\phi,b}
    \end{bmatrix}
    \begin{bmatrix}
        \delta\boldsymbol{\theta} \\
        \delta\boldsymbol{\phi}_b
    \end{bmatrix}.
    \label{eq:local-quadratic}
\end{split}
\end{equation}

The local nuisance increment is then analytically eliminated. With a small numerical damping term $\lambda_{\phi}$ where required, the reduced slow-state information matrix is

\begin{equation}
    \boldsymbol{S}_b
    =
    \boldsymbol{H}_{\theta\theta,b}
    -
    \boldsymbol{H}_{\theta\phi,b}
    \left(
        \boldsymbol{H}_{\phi\phi,b}
        +
        \lambda_{\phi}\boldsymbol{I}
    \right)^{-1}
    \boldsymbol{H}_{\phi\theta,b},
    \label{eq:schur-information}
\end{equation}

and the corresponding reduced information vector is

\begin{equation}
    \boldsymbol{r}_b
    =
    -
    \boldsymbol{g}_{\theta,b}
    +
    \boldsymbol{H}_{\theta\phi,b}
    \left(
        \boldsymbol{H}_{\phi\phi,b}
        +
        \lambda_{\phi}\boldsymbol{I}
    \right)^{-1}
    \boldsymbol{g}_{\phi,b}.
    \label{eq:schur-score}
\end{equation}

The reduced quadratic can therefore be expressed as

\begin{equation}
    \widetilde{\mathcal{L}}_b
    \left(
        \delta\boldsymbol{\theta}
    \right)
    =
    \frac{1}{2}
    \delta\boldsymbol{\theta}^{\mathsf{T}}
    \boldsymbol{S}_b
    \delta\boldsymbol{\theta}
    -
    \boldsymbol{r}_b^{\mathsf{T}}
    \delta\boldsymbol{\theta}
    +
    \mathrm{const.}
    \label{eq:reduced-quadratic}
\end{equation}

At a well-converged local registration solution, $\boldsymbol{g}_{\phi,b}$ is approximately zero, but Eq.~(\ref{eq:schur-score}) retains the general form.

The pair $(\boldsymbol{S}_b,\boldsymbol{r}_b)$ is the Schur-reduced sub-block summary. It answers the operational question: after the registration variables have been allowed to adjust, what information does this image sequence retain about separation, photometry, plate scale, and wavefront state? The summary has dimension set by the slow state rather than by the number of frames, pixels, or local registration variables. This representation preserves the image-domain information while avoiding an impractical observation-wide nuisance-parameter solve. Related local variable-elimination strategies are widely used in bundle adjustment and sliding-window state estimation \cite{Triggs2000,DellaertKaess2017,Demmel2021}.

Within one update window, all accepted sub-block summaries are expressed about the same slow-state reference and use the curvature of the summed image likelihood. Their information can therefore be added consistently with the number of accepted frames and sub-blocks. Summaries from later windows are evaluated after the slow-state reference has changed and are not, in the current implementation, merged into a persistent observation-wide precision matrix.

\subsection{Window-level accumulation and iterative reference updates}
\label{sec:block-accumulation}

Let $k$ index an update window and let $\mathcal{B}_k$ denote its accepted sub-blocks. All summaries in $\mathcal{B}_k$ are evaluated about a common slow-state reference $\boldsymbol{\theta}_{\mathrm{ref},k}$ and can be processed independently. Their window-local reduced information is additive:

\begin{equation}
    \boldsymbol{S}_{k}
    =
    \sum_{b\in\mathcal{B}_k}\boldsymbol{S}_b,
    \qquad
    \boldsymbol{r}_{k}
    =
    \sum_{b\in\mathcal{B}_k}\boldsymbol{r}_b.
    \label{eq:block-accumulation}
\end{equation}

Let the regularizing prior used for window $k$ have mean $\boldsymbol{\theta}_{0,k}$ and precision $\boldsymbol{\Lambda}_{0}$. Expressed about $\boldsymbol{\theta}_{\mathrm{ref},k}$, the window-local posterior precision and information vector are

\begin{equation}
    \boldsymbol{\Lambda}_{k}^{\mathrm{win}}
    =
    \boldsymbol{\Lambda}_{0}
    +
    \boldsymbol{S}_{k},
    \qquad
    \boldsymbol{r}_{k}^{\mathrm{win}}
    =
    \boldsymbol{r}_{k}
    +
    \boldsymbol{\Lambda}_{0}
    \left(
        \boldsymbol{\theta}_{0,k}
        -
        \boldsymbol{\theta}_{\mathrm{ref},k}
    \right).
    \label{eq:posterior-information}
\end{equation}

The proposed physical-basis correction and its local covariance diagnostic are then

\begin{equation}
    \delta\boldsymbol{\theta}_{k}^{\mathrm{post}}
    =
    \left(
        \boldsymbol{\Lambda}_{k}^{\mathrm{win}}
    \right)^{\dagger}
    \boldsymbol{r}_{k}^{\mathrm{win}},
    \qquad
    \boldsymbol{C}_{k}^{\mathrm{win}}
    \simeq
    \left(
        \boldsymbol{\Lambda}_{k}^{\mathrm{win}}
    \right)^{\dagger},
    \label{eq:posterior-update}
\end{equation}

where $\dagger$ denotes the numerically regularized inverse or pseudoinverse. In the current campaigns, the prior mean is re-centered on the updated slow-state reference for each successive window, while the configured prior scales are re-used.

After applying the update policy described in Sec.~\ref{sec:eigenbasis-updates}, the next linearization point is

\begin{equation}
    \boldsymbol{\theta}_{\mathrm{ref},k+1}
    =
    \boldsymbol{\theta}_{\mathrm{ref},k}
    +
    \delta\boldsymbol{\theta}_{k}^{\mathrm{applied}}.
    \label{eq:iterative-reference-update}
\end{equation}

The forward model and local summaries are therefore re-linearized as the retrieval approaches the solution.

The current sequential implementation carries the updated reference state from one window to the next, but it does not carry $\boldsymbol{\Lambda}_{k}^{\mathrm{win}}$ or $\boldsymbol{C}_{k}^{\mathrm{win}}$ forward as an accumulated observation-level belief. Each window instead forms a new local update from that window's Schur summaries and the configured regularizing prior. The final reported separation is the slow-state reference after the last applied update, whereas the reported posterior sigma is derived from the final window's local curvature. It should therefore be interpreted as a final-window information diagnostic, not as the uncertainty obtained by combining all windows in the observation.

A future cumulative estimator could retain earlier likelihood factors or transport their reduced information to the current linearization point before adding the new window. Such a formulation would permit the observation-level uncertainty to contract as independent information is accumulated, but it would require consistent treatment of relinearization, changing parameter couplings, and any evolution allowed in the slow state. The present architecture currently retains only the iteratively updated state between windows.

The formal covariance in Eq.~(\ref{eq:posterior-update}) is consequently both local and model conditioned. Even within one window, it can be miscalibrated when the likelihood is non-quadratic, parameter modes are nearly degenerate, or the truth and inference models differ. ADORA therefore compares the reported posterior sigma against empirical recovered-minus-truth errors across Monte Carlo trials before using it for requirement or performance claims. The high-order-wavefront knowledge-error experiments in Sec.~\ref{sec:results-hoche} provide an explicit test of this distinction by holding the local inference architecture fixed while perturbing a calibrated input to the forward model.

\subsection{Fisher eigenbasis and information-damped updates}
\label{sec:eigenbasis-updates}

The window-local slow-state problem can be strongly anisotropic. In particular, similar low-order aberrations on M1 and M2 produce nearly degenerate changes in the focal-plane image. Separation, plate scale, photometry, and wavefront coefficients also participate in coupled modes. A full physical-basis posterior correction can therefore contain large components along directions that are only weakly constrained by the current window.

ADORA diagnoses and regularizes these directions in an eigenbasis of a window-level information matrix. To compare heterogeneous parameters in dimensionless coordinates, let $\boldsymbol{D}=\operatorname{diag}(\boldsymbol{\sigma}_{0})$ contain the reference prior scales. For a selected information source $\boldsymbol{M}$---for example the data information accumulated within the current window or the corresponding window-local posterior precision---the prior-whitened matrix is

\begin{equation}
    \boldsymbol{G}
    =
    \boldsymbol{D}
    \boldsymbol{M}
    \boldsymbol{D}
    =
    \boldsymbol{V}
    \operatorname{diag}(\gamma_i)
    \boldsymbol{V}^{\mathsf{T}},
    \label{eq:whitened-eigenbasis}
\end{equation}

where $\gamma_i$ measures the information associated with mode $i$ relative to its prior scale. The proposed physical update is transformed into these coordinates,

\begin{equation}
    \boldsymbol{c}_{k}
    =
    \boldsymbol{V}^{\mathsf{T}}
    \boldsymbol{D}^{-1}
    \delta\boldsymbol{\theta}_{k}^{\mathrm{post}}.
    \label{eq:modal-update}
\end{equation}

ADORA then applies a mode-dependent gate $0\leq\eta_i\leq 1$ and an optional global update gain $g$:

\begin{equation}
    \delta\boldsymbol{\theta}_{k}^{\mathrm{applied}}
    =
    g\,
    \boldsymbol{D}
    \boldsymbol{V}
    \operatorname{diag}(\eta_i)
    \boldsymbol{c}_{k}.
    \label{eq:damped-eigen-update}
\end{equation}

For a full eigen update, $\eta_i=1$. In the information-damped policy, $\eta_i$ increases with the information assigned to the mode, so low-information combinations are advanced more cautiously than high-information combinations. A truncated policy instead sets selected poorly constrained modes to zero.

This update is not intended to assert that the eigenvectors are fundamental instrument coordinates. They are local coupled combinations that depend on the target, the current window's pointing history, the reference state, the prior scales, and the information accumulated within that window. 

The value of this formulation is twofold: the eigenbasis stabilizes iterative updates, and they reveal which combinations of science and instrument parameters are actually constrained by a given observation. The eigenvalue spectrum, parameter contributions to each mode, applied damping, and residual physical-basis update are therefore retained as diagnostics. In the current campaigns, this information-aware treatment is particularly important for the strong coupling between corresponding M1 and M2 low-order wavefront modes.

\begin{figure}[htbp]
\centering
\includegraphics[width=\linewidth]{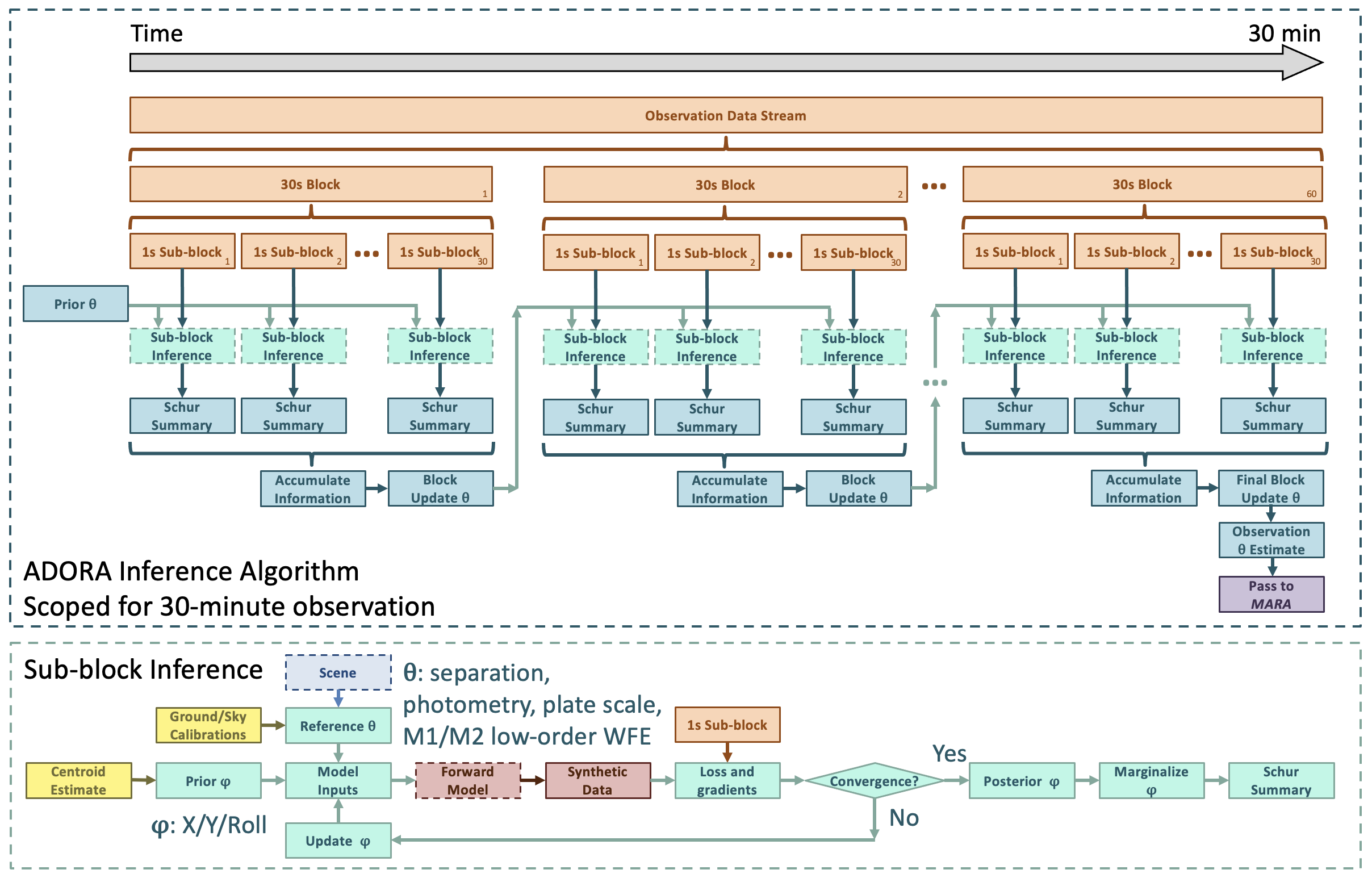}
\caption{ADORA inference algorithm. 
A nominal 30-minute observation is divided into approximately 30-second update blocks and 1-second sub-blocks. Sub-blocks within a block share a slow reference state and are processed independently: the forward model solves the frame X/Y/roll registration, the nuisance variables are eliminated by Schur reduction, and a reduced score and information matrix are retained for the slow state. These summaries are accumulated to form an information-damped window update, which becomes the reference for the next window. The final estimated state is passed to MARA for multi-epoch analysis.
}
\label{fig:adora-retrieval-algorithm}
\end{figure}

\section{Validation Campaign Design}
\label{sec:validation-design}

The validation program compares a data-generating truth model with a separately instantiated inference model. Campaigns retain a common image-formation and retrieval architecture while changing selected initialization, target, duration, or calibration assumptions. This design separates stochastic retrieval behavior from bias caused by fixed truth--inference mismatch.

\subsection{Common truth--inference configuration}
\label{sec:validation-model-split}

For realization $j$, synthetic images are generated from $\mathcal{M}_{j}^{\mathrm{truth}}$ and analyzed with  $\mathcal{M}_{j}^{\mathrm{inf}}$. Variation may enter through the initial slow-state draw and image noise, through recoverable offsets in fitted parameters, or through fixed mismatch in calibrated inputs. Matched-model campaigns disable the controlled wavefront and detector mismatches but retain photon and read noise, time-dependent pointing and smear, low-order initialization errors, and the coupled M1/M2 slow state. Knowledge-error campaigns instead perturb one inference-side calibration while holding it fixed during the observation-level solve, allowing the fitted state to absorb part of the omitted image structure.





The principal baseline uses Alpha Centauri A/B and the three-plane model of Sec. ~\ref{sec:forward-model}. Each trial contains \SI{50}{\milli\second} exposures grouped into 20-frame sub-blocks; 30 sub-blocks form one \SI{30}{\second} update window, and ten windows provide \SI{5}{\minute} of realized data. Two replicated baseline groups draw the fitted M1 and M2 low-order coefficients at characteristic scales of \SI{0.3}{\nano\meter} and \SI{1.0}{\nano\meter}, with ten realizations per group. These are initialization errors, not persistent calibration mismatch.

The production campaigns form Schur summaries using the supplied frame-registration reference when available. Registration variables and their coupling to the slow state remain represented in the reduced curvature, but the expensive local registration solve is not repeated for every production sub-block. The results therefore primarily validate slow-state retrieval and truth--inference mismatch under a supplied pointing solution; recovered-registration behavior has been exercised separately at smaller scale.

\begin{table}[htbp]
\caption{Campaign-specific settings common to the principal full-fidelity validation runs. Model components and fitted parameters are defined in Secs.~\ref{sec:overview}--\ref{sec:retrieval-algorithm}.}
\label{tab:baseline-validation-settings}
\centering
\footnotesize
\begin{tabular}{@{}p{0.25\linewidth}p{0.69\linewidth}@{}}
\toprule
Setting & Common configuration \\
\midrule
Target and duration
& Alpha Centauri A/B; ten \SI{30}{\second} windows (\SI{5}{\minute}) for replicated families. \\
Cadence
& \SI{50}{\milli\second} frames, 20 frames per sub-block, and 30 sub-blocks per update window. \\
Slow and local states
& Separation, flux, contrast, plate scale, and M1/M2 Zernikes fitted; frame X/Y/roll represented as local nuisance variables. \\
Pointing and noise
& Filtered trajectory, matched exposure smear, photon noise, and detector read noise. \\
Calibration split
& Spectra, throughput, high-order WFE, and detector maps matched unless explicitly perturbed in a knowledge-error family. \\
Update policy
& Information-damped eigen update; the reference state, but not an observation-cumulative precision matrix, is propagated between windows. \\
\bottomrule
\end{tabular}
\end{table}

\subsection{Campaign families and reporting conventions}
\label{sec:validation-families}
\label{sec:validation-hoche}
\label{sec:validation-metrics}

The high-order WFE knowledge error (KE) sweep applies inference-side residuals to the high-order WFE maps of both the primary (M1) and secondary (M2) mirrors. For each mirror, the residual has an RMS amplitude of \SI{0.01}{\nano\meter}, \SI{0.1}{\nano\meter}, or \SI{1.0}{\nano\meter}. Five draws are evaluated for each amplitude and low-order initialization regime. The sweep uses a common high-order realization and therefore samples amplitude more extensively than spatial morphology; it is not interpreted as a universal high-order-WFE tolerance curve. The pixel-position sweep similarly uses representative fixed maps at each amplitude; independent detector-map morphologies and pixel-response mismatch are deferred. The 15- and 30-minute matched-model trials contain actual image sequences, but only one realization at each duration.

\begin{table}[htbp]
\caption{Principal validation families. Sample counts are stated per independent condition where applicable.}
\label{tab:validation-families}
\centering
\footnotesize
\begin{tabular}{@{}p{0.20\linewidth}p{0.30\linewidth}p{0.16\linewidth}p{0.26\linewidth}@{}}
\toprule
Family & Controlled variation & Realized sample & Scientific purpose \\
\midrule
Matched model
& Low-order initialization; calibrated inputs matched
& 10 draws per regime, \SI{5}{\minute}
& Establish stochastic recovery and iterative capture. \\
High-order WFE KE
& Static high-order residual applied to both M1 and M2
& 3 amplitudes; 5 draws per amplitude and regime
& Measure mismatch bias and leakage into the fitted state. \\
Pixel-position KE
& Static inference-side detector-geometry error
& 3 amplitudes; 10 draws per amplitude
& Test sensitivity to geometric detector calibration. \\
Target registry sweep
& Target flux, contrast, separation, spectra, and geometry
& 10 draws for each of 7 binaries
& Compare information in a fixed \SI{30}{\second} window. \\
Long-duration pilots
& Number of \SI{30}{\second} windows
& One \SI{15}{\minute} and one \SI{30}{\minute} run
& Demonstrate sustained execution; not duration scaling. \\
\bottomrule
\end{tabular}
\end{table}

For each draw, the signed final separation error is $\Delta\rho=\widehat{\rho}_{\mathrm{final}}-\rho_{\mathrm{true}}$. Group results report the mean signed error as the bias estimator and the sample standard deviation as draw-to-draw scatter; mean and median absolute errors describe achieved residual magnitude rather than bias. The reported $\sigma_{\rho}^{\mathrm{win}}$ is derived from the final window's local curvature and is used as an information and conditioning diagnostic, not as an observation-cumulative uncertainty.

For draw $j$, the signed final separation error is $\Delta\rho_j=\widehat{\rho}_{j,\mathrm{final}}-\rho_{j,\mathrm{true}}$. For a group of $N$ realizations, $\overline{\Delta\rho}$ denotes the mean signed error and is used to estimate ensemble bias, while $s_{\Delta\rho}$ denotes the sample standard deviation of the signed errors and quantifies empirical draw-to-draw scatter about that mean. The standard error on the estimated mean bias is $s_{\Delta\rho}/\sqrt{N}$. Mean and median absolute errors describe achieved residual magnitude rather than bias. The reported $\sigma_{\rho}^{\mathrm{win}}$ is derived from the final window's local curvature and is used as an information and conditioning diagnostic, not as an observation-cumulative uncertainty.

\section{Results}
\label{sec:results}


All headline values below use the final state of the realized observation and the signed-error convention of Sec.~\ref{sec:validation-metrics}.

\subsection{Matched-model recovery and iterative behavior}
\label{sec:results-no-ke-baseline}
\label{sec:results-no-ke-convergence}


Table~\ref{tab:no-ke-baseline-results} summarizes the replicated \SI{5}{\minute} matched-model campaign. The two initialization regimes yield mean signed residuals of $-0.771~\uas$ and $+0.887~\uas$, with sample standard deviations of $10.987~\uas$ and $10.517~\uas$. The corresponding standard errors on the means are $3.474~\uas$ and $3.326~\uas$; no ensemble bias is detected at the current sample depth.

\begin{table}[htbp]
\caption{Matched-model separation recovery after \SI{5}{\minute} of realized data. Astrometric entries are in $\uas$, and $\sigma_{\rho}^{\mathrm{win}}$ is the final-window diagnostic.}
\label{tab:no-ke-baseline-results}
\centering
\footnotesize
\begin{tabular}{@{}lrrrrrr@{}}
\toprule
M1/M2 initialization
& $N$
& $\overline{\Delta\rho}$
& $s_{\Delta\rho}$
& $\overline{|\Delta\rho|}$
& $\operatorname{med}|\Delta\rho|$
& $\overline{\sigma_{\rho}^{\mathrm{win}}}$ \\
\midrule
\SI{0.3}{\nano\meter}
& 10 & $-0.771$ & 10.987 & 8.368 & 5.781 & 17.469 \\
\SI{1.0}{\nano\meter}
& 10 & $+0.887$ & 10.517 & 8.446 & 5.905 & 17.636 \\
\bottomrule
\end{tabular}
\end{table}

Both groups have a mean absolute residual near $8.4~\uas$ and a median absolute residual near $5.8$--$5.9~\uas$, with no evident degradation from the larger initialization. Their final-window posterior sigmas remain $17$--$18~\uas$, somewhat larger than the empirical scatter and consistent with a conservative local curvature diagnostic for this matched-model sample.

\begin{figure}[htbp]
\centering
\includegraphics[width=\linewidth]{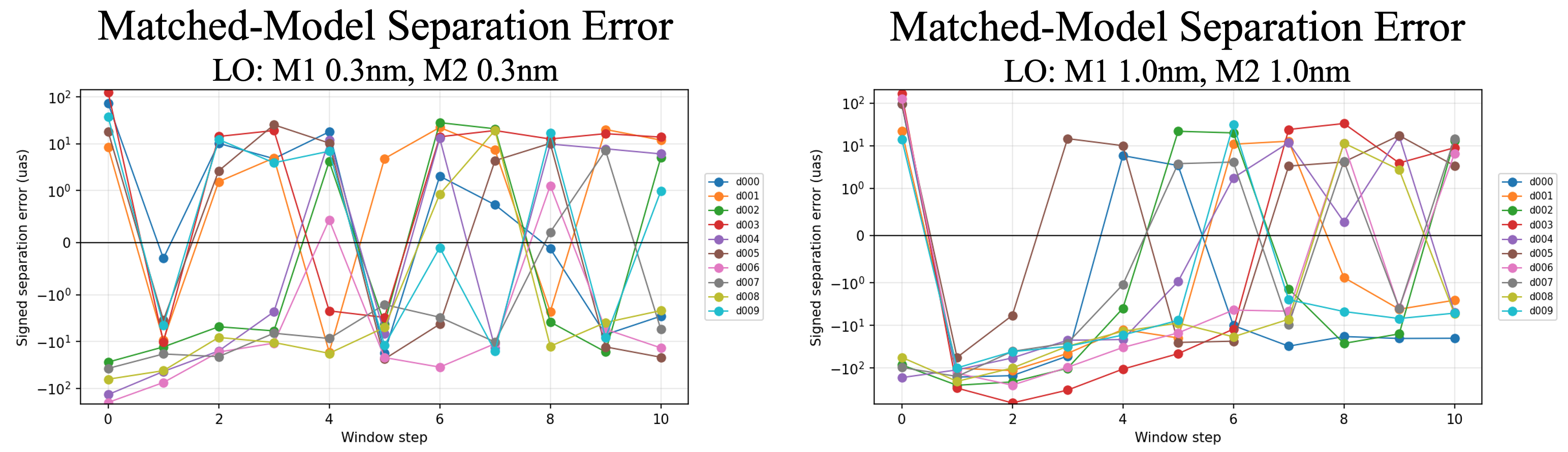}
\caption{Iterative matched-model recovery over ten \SI{30}{\second} update windows. The milder initialization approaches the final error scale quickly, whereas the \SI{1.0}{\nano\meter} initialization exhibits a larger early transient before reaching a similar final separation distribution.}
\label{fig:no-ke-baseline-convergence}
\end{figure}

The trajectories in Fig.~\ref{fig:no-ke-baseline-convergence} show different capture behavior. The \SI{0.3}{\nano\meter} group approaches the final error scale within the first few windows, whereas the \SI{1.0}{\nano\meter} group exhibits a larger nonlinear overshoot before recovering to a statistically similar endpoint distribution. The optical state improves concurrently: the final M1/M2 low-order RMS errors are $0.042/0.105~\mathrm{nm}$ and $0.091/0.207~\mathrm{nm}$ for the two groups, respectively. M2 remains less completely separated from the coupled optical state, even though the recovered separation is similar.

After capture, the updates occupy a roughly stationary, window-conditioned error band rather than showing sustained $1/\sqrt{T}$ contraction, consistent with the non-cumulative architecture of Sec.~\ref{sec:block-accumulation}.


\subsection{High-order wavefront knowledge error}
\label{sec:results-hoche}


Table~\ref{tab:hoche-results} reports the high-order knowledge error endpoints. The \SI{0.01}{\nano\meter} cases remain broadly baseline-like at the present sample depth. At \SI{0.1}{\nano\meter}, the two initialization regimes show positive mean residuals of $28.65~\uas$ and $24.94~\uas$, with all ten endpoints positive. At \SI{1.0}{\nano\meter}, the means increase to $140.96~\uas$ and $142.08~\uas$, again with all ten endpoints positive. The similar large-amplitude results across initialization regimes indicate that the fixed mismatch dominates the final solution for this representative spatial pattern.

\begin{table}[htbp]
\caption{Five-minute separation recovery with inference-side high-order WFE knowledge error. The listed KE amplitude is the RMS residual applied to each of the M1 and M2 high-order WFE maps. Astrometric entries are in $\uas$; $\sigma_{\rho}^{\mathrm{win}}$ remains a model-conditioned final-window diagnostic.}
\label{tab:hoche-results}
\centering
\footnotesize
\begin{tabular}{@{}llrrrrr@{}}
\toprule
HO-KE RMS
& M1/M2 initialization
& $N$
& $\overline{\Delta\rho}$
& $s_{\Delta\rho}$
& $\overline{|\Delta\rho|}$
& $\overline{\sigma_{\rho}^{\mathrm{win}}}$ \\
\midrule
\SI{0.01}{\nano\meter} & \SI{0.3}{\nano\meter}
& 5 & $+1.65$ & 6.99 & 5.19 & 17.47 \\
\SI{0.01}{\nano\meter} & \SI{1.0}{\nano\meter}
& 5 & $-4.12$ & 18.25 & 13.82 & 17.64 \\
\SI{0.1}{\nano\meter} & \SI{0.3}{\nano\meter}
& 5 & $+28.65$ & 12.80 & 28.65 & 17.47 \\
\SI{0.1}{\nano\meter} & \SI{1.0}{\nano\meter}
& 5 & $+24.94$ & 8.78 & 24.94 & 17.64 \\
\SI{1.0}{\nano\meter} & \SI{0.3}{\nano\meter}
& 5 & $+140.96$ & 16.55 & 140.96 & 17.49 \\
\SI{1.0}{\nano\meter} & \SI{1.0}{\nano\meter}
& 5 & $+142.08$ & 24.80 & 142.08 & 17.73 \\
\bottomrule
\end{tabular}
\end{table}

The fitted state increasingly absorbs the omitted image structure. At \SI{1.0}{\nano\meter} RMS, the final low-order coefficient RMS errors reach approximately $1.84$--$2.00~\mathrm{nm}$ on M1 and $4.21$--$4.60~\mathrm{nm}$ on M2, while plate scale and source parameters also shift. The final-window separation sigma nevertheless remains $17$--$18~\uas$ across the sweep, leaving the \SI{1.0}{\nano\meter} solutions about eight local sigmas from truth. The larger-amplitude trajectories are therefore biased pseudo-solutions rather than a simple increase in zero-mean measurement scatter.

\begin{figure}[htbp]
\centering
\includegraphics[width=\linewidth]{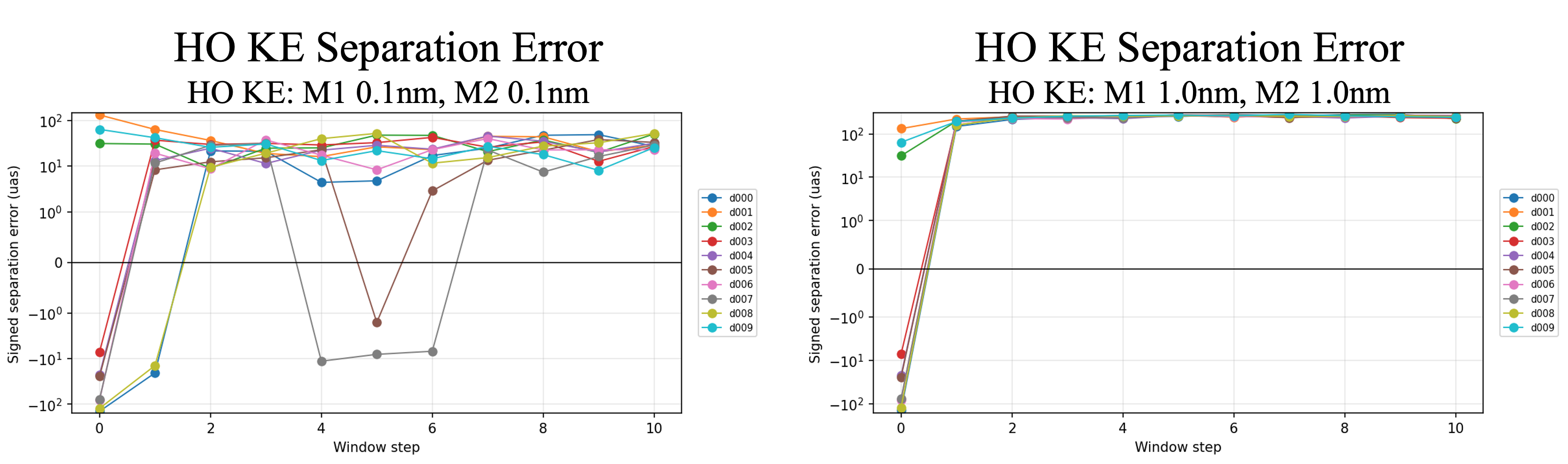}
\caption{Signed separation error for the \SI{0.1}{\nano\meter} (left) and \SI{1.0}{\nano\meter} (right) high-order WFE knowledge-error conditions. The larger mismatches drive the retrieval toward a repeatable positive separation bias.}
\label{fig:hoche-sensitivity}
\end{figure}

\subsection{Pixel-position knowledge error}
\label{sec:results-pixel-position-ke}


\begin{table}[htbp]
\caption{Five-minute separation recovery with inference-side pixel-position knowledge error. The matched row uses the same \SI{0.3}{\nano\meter} initialization regime. Astrometric entries are in $\uas$.}
\label{tab:pixel-position-ke-results}
\centering
\footnotesize
\begin{tabular}{@{}lrrrrr@{}}
\toprule
Pixel-position KE
& $N$
& $\overline{\Delta\rho}$
& $s_{\Delta\rho}$
& $\overline{|\Delta\rho|}$
& $\overline{\sigma_{\rho}^{\mathrm{win}}}$ \\
\midrule
Matched No-KE
& 10 & $-0.77$ & 10.99 & 8.37 & 17.47 \\
$10^{-4}$ pixel
& 10 & $+2.17$ & 12.97 & 8.53 & 17.05 \\
$5\times10^{-4}$ pixel
& 10 & $-2.30$ & 17.28 & 14.15 & 17.05 \\
$10^{-3}$ pixel
& 10 & $-9.46$ & 11.28 & 11.04 & 17.05 \\
\bottomrule
\end{tabular}
\end{table}



Across $10^{-4}$--$10^{-3}$ pixel, all conditions recover away from their initial draws and finish with $11$--$17~\uas$ empirical scatter, while the final-window sigma remains $17.05~\uas$. The $10^{-4}$-pixel case is baseline-like; the higher amplitudes show modest but non-monotonic changes in bias and scatter. The present sweep therefore supports robustness to the representative position maps tested, but not a detector-position scaling law or formal calibration requirement.

\subsection{Target-dependent window information and duration pilots}
\label{sec:results-target-information}
\label{sec:results-long-no-ke}


The target-registry sweep applies the same approximately \SI{30}{\second} update window to seven binary systems. Table~\ref{tab:target-window-information} reports the mean final-window posterior separation sigma for each target, averaged over ten realizations. These values quantify the local, model-conditioned information available within one representative \SI{30}{\second} update window; they are not cumulative observation-level uncertainties.

\begin{table}[ht]
    \centering
    \caption{Target-dependent separation information for a representative approximately \SI{30}{\second} update window. The reported values are the mean final-window posterior separation sigmas over ten realizations per target. They characterize local statistical information under the assumed model and prior, rather than achieved five-minute errors or cumulative 30-minute uncertainties.}
    \label{tab:target-window-information}
    \begin{tabular}{lc}
        \hline
        Binary system & Mean window-local $\sigma_{\rho}^{\rm win}$ ($\uas$) \\
        \hline
        Alpha Centauri   & 17.64 \\
        36 Ophiuchi      & 73.67 \\
        70 Ophiuchi      & 77.24 \\
        $\xi$ Bootis     & 86.22 \\
        p Eridani        & 92.15 \\
        HR 2667/2668     & 97.83 \\
        61 Cygni         & 99.19 \\
        \hline
    \end{tabular}
\end{table}

The approximately five-fold range in window-local sigma demonstrates that equal elapsed time does not provide equal astrometric information across the target set. Target flux is likely a major contributor, although contrast, nominal separation, component spectra, and overlap of the structured PSFs also vary simultaneously and are not isolated by the present registry campaign. Because the update cadence is scoped for Alpha Centauri, the currently untuned endpoint convergence is not representative of intrinsic target performance.

Under the idealized assumption that statistically equivalent and independent window information accumulates with the usual $1/\sqrt{T}$ scaling, the Alpha Centauri value would correspond to $17.64/\sqrt{60}\simeq2.28~\uas$ over a \SI{30}{\minute} observation. Applying the same illustrative scaling gives approximately $9.51$--$12.81~\uas$ for the other six targets. These values can be interpreted as statistical floors rather than predictions of achieved 30-minute performance: the present posterior sigma includes the configured prior, the information content may vary between windows, and the current estimator does not yet retain a cumulative observation-wide precision matrix.

Two single-realization matched-model pilots extended the same window definition to actual durations of \SI{15}{\minute} and \SI{30}{\minute}. Their final residuals were $-8.225~\uas$ and $-0.025~\uas$, while their final-window sigmas remained $16.771~\uas$ and $17.075~\uas$. Both runs demonstrate sustained execution, but neither the favorable 30-minute endpoint nor the two individual trajectories establish duration scaling or a cumulative observation-level precision.





\section{Discussion and outlook}
\label{sec:discussion}

The campaign suite establishes a matched-model recovery baseline and demonstrates that ADORA can diagnose how statistical information and fixed calibration mismatch enter the same image-domain retrieval. The results remain an initial validation rather than a complete mission-performance assessment, but they identify the estimator developments and calibration studies most likely to affect SHERA astrometry.


\subsection{Window-local information, cadence, and cumulative estimation}
\label{sec:discussion-information-accumulation}
\label{sec:discussion-adaptive-cadence}

The matched-model trajectories show successful capture from substantially different low-order initializations, followed by fluctuations within a roughly stationary error band. This behavior is consistent with the present estimator carrying the updated reference state, but not the full posterior precision, from one window to the next. The repeated $17$--$18~\uas$ posterior sigma should therefore be interpreted as the local curvature of one approximately \SI{30}{\second} block under the assumed model, not as the uncertainty of a joint five- or thirty-minute posterior.

The binary target sweep shows that this window-local information is strongly scene dependent: a fixed \SI{30}{\second} block is much more informative for Alpha Centauri than for the six other binaries tested. A single elapsed-time cadence is consequently unlikely to be optimal across the mission target set. Future updates could accumulate data until selected modes reach an information threshold, delay weakly constrained modal corrections, or vary the window length with target and convergence state while respecting the timescale over which the slow state remains approximately constant.

Observation-cumulative estimation is complementary to cadence adaptation. Retaining earlier reduced likelihood factors, propagating an information-form belief, or periodically relinearizing accumulated factors could permit uncertainty to contract as data accumulates. Because the forward model is nonlinear and successive summaries are formed at different reference states, this is more involved than summing historical Hessians; a retained-factor, incremental-smoothing, or fixed-lag formulation must account for relinearization and any allowed slow-state evolution. Additional campaigns are then needed to test genuine integration-time scaling.

\subsection{Calibration-model mismatch}
\label{sec:discussion-calibration-coupling}
\label{sec:discussion-pixel-position-ke}

The high-order and pixel-position knowledge error families illustrate two different calibration regimes. High-order WFE mismatch projects strongly into separation and the fitted optical and source state, producing a locally well-conditioned but physically biased solution. Additional data processed with the same mismatch is expected to converge towards the same biased offset.

By contrast, the tested pixel-position knowledge errors remain near the baseline matched-model retrieval scale. This does not imply that detector geometry is unimportant; it shows that, for the representative maps and amplitudes tested, it couples less strongly into separation than the high-order WFE residual. Different calibration errors may lead to different astrometric consequences even when they leave the local Fisher information nearly unchanged.

Neither sweep yet defines a formal requirement. The high-order knowledge-error study varies amplitude more extensively than morphology, and the pixel-position study uses representative fixed maps rather than an ensemble of independent realizations. Requirement-level flowdown will require a more careful study of different knowledge-error realizations, denser sampling near sensitivity transitions, dedicated pixel-response sweeps, and interactions with fitted parameters. Calibration-model uncertainty may ultimately need to be propagated using ensembles of plausible calibration states, estimated jointly with the astrometric state when the relevant calibration parameters are identifiable, or represented through an explicit model-discrepancy term.

\subsection{Limitations and validation priorities}
\label{sec:discussion-limitations}



The principal limitations are the modest ensemble sizes, single-realization long-duration trials, supplied registration states in the large production campaigns, window-local rather than cumulative uncertainty propagation, and incomplete coverage of spectral, pointing, and combined calibration errors. The current results validate components of the architecture and identify controlled sensitivities; they do not yet constitute a complete SHERA astrometric error budget or final mission-performance prediction.

The next validation priorities are:
\begin{enumerate}
    \item expand optical and detector knowledge-error studies across independent map realizations, intermediate amplitudes, pixel-response detector mismatch, and spectral calibration error;
    \item implement and test observation-cumulative information propagation together with information-gated or target-adaptive update cadence;
    \item repeat campaigns with sufficient Monte Carlo depth and evaluate production-scale recovered registration; and
    \item combine multiple representative calibration errors in an integrated end-to-end ensemble after the individual sensitivities are understood.
\end{enumerate}

As these studies mature, the same differentiable forward model and Schur-summary infrastructure can compare observing strategies, calibration assumptions, and update policies on a common image-domain basis. The present campaigns provide both an initial validation of ADORA and a diagnostic framework for identifying instrument errors that can masquerade as astrophysical motion.


\acknowledgments 
 
The research was carried out at the Jet Propulsion Laboratory, California Institute of Technology, under a contract with the National Aeronautics and Space Administration (80NM0018D0004).

\copyright 2026. All rights reserved.

\bibliography{report} 

@incollection{Triggs2000,
  author    = {Triggs, Bill and McLauchlan, Philip F. and Hartley, Richard I. and Fitzgibbon, Andrew W.},
  title     = {Bundle Adjustment---A Modern Synthesis},
  booktitle = {Vision Algorithms: Theory and Practice},
  editor    = {Triggs, Bill and Zisserman, Andrew and Szeliski, Richard},
  series    = {Lecture Notes in Computer Science},
  volume    = {1883},
  year      = {2000},
  note      = {Proceedings of IWVA 1999}
}

@inproceedings{Demmel2021,
  author    = {Demmel, Nikolaus and Schubert, David and Sommer, Christian and Cremers, Daniel and Usenko, Vladyslav},
  title     = {Square Root Marginalization for Sliding-Window Bundle Adjustment},
  booktitle = {2021 IEEE/CVF International Conference on Computer Vision (ICCV)},
  address   = {Montreal, QC, Canada},
  pages     = {13240--13248},
  year      = {2021},
  doi       = {10.1109/ICCV48922.2021.01301}
}

@article{DellaertKaess2017,
  author  = {Dellaert, Frank and Kaess, Michael},
  title   = {Factor Graphs for Robot Perception},
  journal = {Foundations and Trends in Robotics},
  volume  = {6},
  pages   = {1--139},
  year    = {2017}
}

@article{Desdoigts2023,
  author  = {Desdoigts, Louis and Pope, Benjamin J. S. and Dennis, J. and Tuthill, Peter G.},
  title   = {Differentiable Optics with {$\partial$Lux}: I---Deep Calibration of Flat Field and Phase Retrieval with Automatic Differentiation},
  journal = {Journal of Astronomical Telescopes, Instruments, and Systems},
  volume  = {9},
  number  = {2},
  pages   = {028007},
  year    = {2023},
  doi     = {10.1117/1.JATIS.9.2.028007}
}

@misc{Desdoigts2024,
  author        = {Desdoigts, Louis and Pope, Benjamin and Gully-Santiago, Michael and Tuthill, Peter},
  title         = {Differentiable Optics with {dLux} II: Optical Design Maximising Fisher Information},
  year          = {2024},
  eprint        = {2406.08704},
  archiveprefix = {arXiv},
  doi           = {10.48550/arXiv.2406.08704},
  note          = {Preprint; publication status to be verified before submission}
}

@article{Desdoigts2026,
  author  = {Desdoigts, Louis and others},
  title   = {{AMIGO}: A Data-Driven Calibration of the {JWST} Interferometer},
  journal = {Publications of the Astronomical Society of Australia},
  pages   = {1--37},
  year    = {2026},
  doi     = {10.1017/pasa.2026.10194},
  note    = {Author list, volume, issue, and article number to be verified before submission}
}

@article{Lindegren2012,
  author  = {Lindegren, Lennart and Lammers, Uwe and Hobbs, David and O'Mullane, William and Bastian, Ulrich and Hern{\'a}ndez, Josefa},
  title   = {The Astrometric Core Solution for the {Gaia} Mission: Overview of Models, Algorithms, and Software Implementation},
  journal = {Astronomy \& Astrophysics},
  volume  = {538},
  pages   = {A78},
  year    = {2012},
  doi     = {10.1051/0004-6361/201117905}
}

@unpublished{Christiansen2026,
  author = {Christiansen, Jessie L. and Mamajek, Eric E. and Meyer, Michael R. and others},
  title  = {Searching for Habitable Exoplanets with Relative Astrometry
            ({SHERA}). I. The Case for Searching for Planets in Binary Star
            Systems},
  year   = {2026},
  note   = {Manuscript in preparation; author list and final bibliographic
            details to be updated before submission}
}

@unpublished{Roberson2026,
  author = {Roberson, Bill and others},
  title  = {Orbitize! for SHERA},
  year   = {2026},
  note   = {Manuscript in preparation; author list and final bibliographic
            details to be updated before submission}
}

@article{Shao1992,
  author  = {Shao, Michael and Colavita, Mark M.},
  title   = {Potential of Long-Baseline Infrared Interferometry for Narrow-Angle
             Astrometry},
  journal = {Astronomy \& Astrophysics},
  volume  = {262},
  pages   = {353--358},
  year    = {1992}
}

@article{Lane2004,
  author  = {Lane, Benjamin F. and Muterspaugh, Matthew W.},
  title   = {Differential Astrometry of Subarcsecond Scale Binaries at the
             {P}alomar Testbed Interferometer},
  journal = {The Astrophysical Journal},
  volume  = {601},
  pages   = {1129--1135},
  year    = {2004},
  doi     = {10.1086/380760}
}
\bibliographystyle{spiebib} 

\end{document}